\documentclass[prb,aps,twocolumn,amsmath,amssymb,superscriptaddress,color,showpacs]{revtex4}
\usepackage{latexsym}
\usepackage{amsmath}
\usepackage{amssymb}
\usepackage{graphicx}
\usepackage{caption}
\usepackage{subfigure}
\usepackage{float}
\usepackage{mathrsfs}
\usepackage{color}

\begin{document}
\title{Nonequilibrium-induced enhancement of dynamical quantum coherence and entanglement of spin arrays}

\author{Zhedong Zhang}
\affiliation{Department of Physics and Astronomy, SUNY Stony Brook, Stony Brook, NY 11794, USA}
\affiliation{Department of Chemistry, University of California Irvine, Irvine, CA 92697, USA}
\author{Hongchen Fu}
\affiliation{School of Physics and Energy, Shenzhen University, Shenzhen 518060, China}
\author{Jin Wang}
\email{jin.wang.1@stonybrook.edu}
\affiliation{Department of Physics and Astronomy, SUNY Stony Brook, Stony Brook, NY 11794, USA}
\affiliation{Department of Chemistry, SUNY Stony Brook, Stony Brook, NY 11794, USA}
\affiliation{State Key Laboratory of Electroanalytical Chemistry, Changchun Institute for Applied Chemistry, Chinese Academy of Sciences, Changchun, Jilin 130022, China}
\date{\today}

\begin{abstract}
The random magnetic field produced by nuclear spins has long been viewed as the dominating source of decoherence in the quantum-dot based spins. Here we obtain in both {\it exact} and {\it analytical} manner the dynamics of spin qubits coupled to nuclear spin environments via the hyperfine interaction, going beyond the weak system-bath interaction and Markovian approximation. We predict that the detailed-balance-breaking produced by chemical potential gradient in nuclear baths leads to the rapid oscillations of populations, quantum coherence and entanglement, which are absent in the conventional case (i.e., Overhauser noise). This is attributed to the nonequilibruim feature of the system as shown in the relation between the oscillation period and the chemical potential imbalance. Our results reveal the essentiality of nonequilibriumness with detailed-balance-breaking for enhancing the dynamical coherence and entanglement of spin qubits. Moreover our exact solution explicitly demonstrates that the non-Markovian bath comprised by nuclear spins can preserve the collective quantum state, due to the recovery of coherence. Finally we propose an experiment using ultracold trapped ions to observe these nonequilibrium and memory effects.
\end{abstract}

\maketitle

\section{Introduction}
Recently the quantum-dot-based spin qubits as a solid-state method are demonstrated to be successful in quantum information science, such as the control of electron-spin qubits in GaAs quantum dots \cite{Yamamoto08,Koppens05,Yacoby09,Gossard09} and sensitive metrology \cite{Yacoby12}. This takes the advantage of the controllability of the systems. Despite the controllability, the decoherence owing to the influence of nuclei spins in the host materials still remains challenging for maintaining a high fidelity during the quantum computing. Understanding the dynamics of this process is of fundamental importance.

In the materials the spins are subjected to the noise, due to the random magnetic field from the nuclear spins. This is so-called Overhauser field under typical conditions, producing random magnitudes and directions. 
This practically results in the decoherence in a typical timescale. So far, the dynamics of spin arrays has been mostly explored under the influence of Overhauser fields, through the considerable studies on how to mitigate the decoherence or alternatively electron-spin flip \cite{Petta05,Yacoby11,Levy02} and dynamical decoupling \cite{Uhrig07,Sarma07} to artificially eliminate the random noise.  These studies often assume that the spins will eventually relax to the equilibrium state. However, the natural environments as the nuclei spins around GaAs or Si quantum dots always show the emergence of the inhomogenous charge density, which gives the gradient of chemical potentials. Regardless of the polarization fluctuation, the inhomogeneity here refers to the effect of magnitude randomness of the nuclei spins, which is dictated by the charge density associated with nuclei spins. This generates an effective voltage, leading to the breakdown of detailed-balance (time-reversal symmetry). Hence the system will relax to a nonequilibrium steady state breaking the time-reversal-symmetry, rather than being thermalized to the equilibrium. Inspired by these, it is necessary to consider the issue from a new point of view of nonequilibrium quantum dynamics with detailed-balance-breaking, although certain properties can still be explored in the conventional framework \cite{Deng16} under the equilibrium idea. Recently many interesting phenomenons, i.e., the improvement of stationary coherence and current \cite{Esposito06,Zhang14} in quantum dots, quantum synchronization with robust phase-locking \cite{Holland2014_PRL,Walter2014_PRL,Ludwig2013_PRL}, were found to be inherently nonequilibrium. This clearly shows the inadequancy of the conventional ideas. Hence for the dynamical processes the new insight from the nonequilibriumness should be adopted for further profound understanding of the system relaxations. For instance, the propagation of the spin wave associated with the spin transfer in quantum dots would show certain remarkable behavior thanks to the nonequilibriumness induced by inhomogeneity of charge density. This is what we will focus in this article. 

We uncover a novel phase, which shows the rapid oscillation of coherence, fidelity of the collective quantum states of spin qubits and quantum entanglement, contrary to the pure decay under the conventional equilibrium condition. This, in other words, reveals that the nonequilibriumness is essential for enhancing the dynamical coherence of the systems, besides steady-state coherence \cite{Zhang14,Zhang15pccp,Sun15}. This coherent-oscillation phase originates intrinsically from the nonequilibrium-induced net current, which quantifies the degree of deviation from equilibrium. Compared to the previous work, our {\it exact} and {\it analytical} solution to the dynamics of spin qubits in the presence of noisy environments goes beyond the weak system-bath coupling and Markovian approximations \cite{Esposito06,Schuetz12,Schuetz16}. This demonstrates in a general scenario the revival of quantum coherence arising from the non-Markovian effect, supported by the pure numerical simulations \cite{Deng16}. We use ultracold trapped ions in the spirit of quantum simulation to propose a detailed experiment for observing this effect and to predict the nonequilibrium-induced coherent oscillation can be seen in the system with as few as two ions, which is accessible in the present experiments \cite{Wunderlich09,Islam11,Duan11}.

\section{Model}
We consider the arrays of two quantum dots where each contains one electron spin (qubit) subject to its own random magnetic field produced by nuclear spins via hyperfine interactions. The inhomogenous charge density of nuclei spins leads to the chemical potential imbalance in the environment around the quantum dots, regardless of the dimensional and geometric details. Our purpose  is to study the nonequilibrium effect induced by this inhomogeneity of the nuclei environment in the host materials (i.e., GaAs or Si). To this end, the disorder of nuclear spins in different dimensions as well as geometry (i.e., nuclear isotopes in 3D volume) does not play a significant role so that it will be considered no longer here. To capture the feature of inhomogenous charge density, we assume that the nucleis in the host materials are described by one-dimensional spin chains with different chemical potentials and the nuclear spins are equally spaced by $a$ in each chain. Thus the original Hamiltonian takes the Heisenberg form of
\begin{equation}
\begin{split}
H = -J{\boldsymbol\sigma}_1\cdot{\boldsymbol\sigma}_2 - \sum_{i=1}^2\sum_{n=1}^N t\ \textbf{S}_{n,i}\cdot\textbf{S}_{n+1,i} + \sum_{i=1}^2\sum_{n=1}^N f{\boldsymbol\sigma}_i\cdot\textbf{S}_{n,i}
\end{split}
\label{ht}
\end{equation}
where $J,\ t$ are the spin-spin coupling strengths for electron and nuclear spins, respectively. In our model, nuclear and qubit spins are modeled as XY model including the transverse coupling only in order to generate the nuclear spin flips \cite{Loss2006_PRB,Urbaszek2013_RMP} and electronic spin-state transitions \cite{Urbaszek2013_RMP}. The longitudinal terms $S_{n,i}^z S_{n,i+1}^z,\ \sigma_1^z\sigma_2^z$ are neglected here because of no transition induced. For the consideration of noise produced by nuclear spins, we only take into account of the magnitude randomness of nuclear spins and neglect the fluctuation of their orientation. Thus the longitudinal component of the random magnetic field generated by nuclear spins dominates and the qubit-nuclear interaction is encoded as Ising type $\sigma_i^z S_{n,i}^z$. This approximation can reasonably apply for the mesoscopic spin system formed by many nuclear spins in a semiconductor quantum dot, due to the fact that the nuclear spin flips are of the off-resonant frequency to the electronic ones especially at strong magnetic fields \cite{Urbaszek2013_RMP,Dial13}. To this end, we will work under the following effective Hamiltonian
\begin{equation}
\begin{split}
H_{\text{eff}} = & -J\left(\sigma_1^x\sigma_2^x+\sigma_1^y\sigma_2^y\right) + \sum_{i=1}^2\sum_{n=1}^N f\sigma_i^z S_{n,i}^z\\[0.15cm]
& \qquad -\sum_{i=1}^2\sum_{n=1}^N t\left(S_{n,i}^x S_{n+1,i}^x+S_{n,i}^y S_{n+1,i}^y\right) 
\end{split}
\label{heff}
\end{equation}
 $\sigma_1^{\pm}=\sigma^{\pm}\otimes 1,\ \sigma_2^{\pm}=\sigma^z\otimes\sigma^{\pm}$ and $S_{n,i}^{\pm}=S^z\otimes S^z\otimes\cdots\otimes S^z\otimes S^{\pm}\otimes 1\otimes\cdots\otimes 1$, based on the mapping between XXZ spin-$\frac{1}{2}$ chain and 1D Fermi-Hubbard model \cite{Rigol11}, and $\sigma^{\pm}=\frac{1}{2}(\sigma^x\mp i\sigma^y),\ S^{\pm}=\frac{1}{2}(S^x\mp iS^y)$ are the standard Pauli matrices. Therefore $\sigma_n^{\pm},\ S_{n,i}^{\pm}$ obey the fermionic anti-commutation relationship.
 
\section{Time evolution of the density matrix of spin qubits}
To solve the dynamics of the entire system, the following collective operators $I_i^{\pm}=\sum_{j=1}^2 O_{ij}\sigma_j^{\pm}$, namely $I_1^{\pm}=\frac{1}{\sqrt{2}}(-\sigma_1^{\pm}+\sigma_2^{\pm}),\ I_2^{\pm}=\frac{1}{\sqrt{2}}(\sigma_1^{\pm}+\sigma_2^{\pm})$ and
\begin{equation}
\begin{split}
& S_{k,i}^{\pm} = \frac{1}{\sqrt{N}}\sum_{n=1}^N S_{n,i}^{\pm}e^{inka},\quad k=\left(\frac{2m}{N}-1\right)\frac{\pi}{a};\\[0.15cm]
& m=1,2,\cdots,N
\end{split}
\label{collective}
\end{equation}
in momentum space, will be firstly introduced to diagonalize the free Hamiltonian. The time evolution of entire system is governed by the operator ${\cal U}(t)=\text{exp}\left[-\frac{i}{\hbar}\int_0^t V_{int}(\tau)d\tau\right]$ where $V_{int}(t)=V_{int}^{(1)}(t)+V_{int}^{(2)}(t)$ in the interaction picture
\begin{equation}
\begin{split}
V_{int}^{(n)}(t) = & f\sum_{i,j=1}^2\sum_k O_{ni}^{-1}O_{nj}^{-1}e^{-i(\omega_i-\omega_j)t}\\[0.15cm]
& \qquad \times\left(I_i^-I_j^+-I_j^+I_i^-\right)\otimes S_{k,n}^z
\end{split}
\label{hint}
\end{equation}
In order to obtain the dynamics of the system, one notices that the operators $l=\frac{1}{2}(I_1^-I_1^+-I_1^+I_1^-+I_2^-I_2^+-I_2^+I_2^-),\ \eta_3= \frac{1}{2}(I_1^-I_1^+-I_1^+I_1^--I_2^-I_2^++I_2^+I_2^-),\ \eta_1=I_1^-I_2^+ + I_2^-I_1^+,\ \eta_2=-i(I_1^-I_2^+ - I_2^-I_1^+)$
satisfy $[l,\eta_j]=0,\ [\eta_i,\eta_j]=2i\epsilon_{ijk}\eta_k$, which gives rise to the Lie algebra $su(2)\oplus u(1)$. Thereby the qubit-nuclear interaction in Eq.(\ref{hint}) does transform according to the irreducible representation $\mathcal{D}_{0}\oplus\mathcal{D}_0\oplus\mathcal{D}_{\frac{1}{2}}$ of the Lie group $SU(2)\times U(1)$. As inspired by the recent experiments \cite{Yacoby11,Dial13} the spin qubits are initially engineered at the state $|S\rangle=\text{cos}\frac{\phi}{2}|1\rangle\otimes|0\rangle+e^{i\theta}\text{sin}\frac{\phi}{2}|0\rangle\otimes|1\rangle$, which is the eigenstate of $l$ with the eigenvalue of 0. This further means the dynamical evolution of qubits according to the irreducible representation $\mathcal{D}_{\frac{1}{2}}$ of group $SU(2)\times U(1)$. $\eta_j$ take the form of $2\times 2$ representation
\begin{equation}
\begin{split}
\eta_1=\begin{pmatrix}
        0 & -1\\[0.1cm]
        -1 & 0\\
       \end{pmatrix},\quad
\eta_2=\begin{pmatrix}
        0 & -i\\[0.1cm]
        i & 0\\
       \end{pmatrix},\quad
\eta_3=\begin{pmatrix}
        -1 & 0\\[0.1cm]
        0 & 1\\
       \end{pmatrix}
\end{split}
\label{eta}
\end{equation}
Hence in terms of $l$ and $\eta_j$, the representation of the time-evolution operator can be found
\begin{equation}
\begin{split}
{\cal U}(t) & = e^{-\frac{i}{\hbar}\int_0^t V_{int}^{(1)}(\tau)d\tau}e^{-\frac{i}{\hbar}\int_0^t V_{int}^{(2)}(\tau)d\tau}\\[0.15cm] 
            & = A_N + iB_N\left(\sigma_x\text{Re}(p)-\sigma_y\text{Im}(p)\right)
\end{split}
\label{evolution}
\end{equation}
by some algebra, where $\beta\equiv \omega t=4Jt/\hbar$ and $C_v\equiv S_{v,1}^z S_{v,2}^z,\ D_v\equiv S_{v,1}^z-S_{v,2}^z$. 
\begin{equation}
\begin{split}
& A_N = \frac{1}{2}\left[\prod_{v=1}^N\left(a+C_v b+i|p|D_v\right)+\text{h.c.}\right]\\[0.15cm]
& B_N = \frac{1}{2i|p|}\left[\prod_{v=1}^N\left(a+C_v b+i|p|D_v\right)-\text{h.c.}\right]
\end{split}
\label{AB}
\end{equation}
and
\begin{equation}
\begin{split}
& a\equiv \text{cos}^2\left(\frac{f}{2J}\text{sin}\frac{\beta}{2}\right),\quad b\equiv \text{sin}^2\left(\frac{f}{2J}\text{sin}\frac{\beta}{2}\right)\\[0.15cm]
& p\equiv \frac{1}{2}e^{i\frac{\beta}{2}}\text{sin}\left(\frac{f}{J}\text{sin}\frac{\beta}{2}\right)
\end{split}
\label{abc}
\end{equation}

In many circumstances, however, the dynamics of spin qubits is what people are mostly interested. For this purpose we further trace out the degrees of freedoms of nuclear spins in the entire density matrix $\rho(t)$ to obtain the non-unitary time evolution of spin qubits. Suppose that initially the qubits are engineered at the state
\begin{equation}
\begin{split}
|\Psi(0)\rangle=\frac{1}{\sqrt{2}}(|1\rangle\otimes |0\rangle+e^{-i\theta}|0\rangle\otimes |1\rangle)
\end{split}
\label{psi0}
\end{equation}
and the nuclear spins are at thermal equilibrium with different chemical potentials (Fermi energies). In other words, the initial condition takes the product form
\begin{equation}
\begin{split}
& \rho(0) = |\Psi(0)\rangle\langle\Psi(0)|\otimes\rho_{nucl}^{\mu_1}\otimes\rho_{nucl}^{\mu_2}\\[0.15cm] 
& \rho_{nucl}^{\mu_i} = Z_i^{-1}e^{-\bar{\beta}(H_{nucl}^{(i)}-\mu_i N_i)};\\[0.15cm]
& H_{nucl} = -t\sum_{i=1}^2\sum_{n=1}^N \textbf{S}_{n,i}\cdot\textbf{S}_{n+1,i};\ i=1,2
\end{split}
\label{rho0}
\end{equation}
where $N_i=\sum_{n=1}^N S_{n,i}^+ S_{n,i}^-$ is the operator of total number of spin excitation in the nuclear spin chain.

After a lengthy but straightforward calculation one can reach the $2\times 2$ density matrix of spin qubits at moment $t$, by tracing out the nuclear spins as environments:
\begin{equation}
\begin{split}
\rho_s(t)=\text{Tr}_B [\rho(t)]=\begin{pmatrix}
                                 \rho_{10,10}(t) & \rho_{10,01}(t)\\[0.15cm]
                                 \rho_{10,01}^*(t) & \rho_{01,01}(t)\\
                                \end{pmatrix}
\end{split}
\label{rhos}
\end{equation}
where $\rho_{10,10}(\rho_{01,01})$ is the population on the state $|1,0\rangle(|0,1\rangle)$ and $\rho_{10,01}$ is the coherence between these two states. They can be written in the analytical form
\begin{widetext}
\begin{equation}
\begin{split}
& \text{Re}\rho_{10,01}(t) = \left(\prod_{v=1}^N K_v + \mathcal{F}_e(\mu_1,\mu_2,t)\right)\frac{\text{cos}\theta}{2} +  \mathcal{F}_o\left(\mu_1,\mu_2,t\right)\text{sin}\theta\text{cos}\frac{\beta}{2}\\[0.2cm]
& \text{Im}\rho_{10,01}(t) = \left(\text{cos}^2\frac{\beta}{2}\prod_{v=1}^N K_v-\text{sin}^2\frac{\beta}{2}+\mathcal{F}_e(\mu_1,\mu_2,t)\text{cos}^2\frac{\beta}{2}\right)\frac{\text{sin}\theta}{2} - \mathcal{F}_o\left(\mu_1,\mu_2,t\right)\text{cos}\theta\text{cos}\frac{\beta}{2}\\[0.2cm]
& \rho_{10,10}(t)-\rho_{01,01}(t) =  \left(1+\prod_{v=1}^N K_v+\mathcal{F}_e(\mu_1,\mu_2,t)\right)\frac{\text{sin}\theta}{2}\text{sin}\beta - 2\mathcal{F}_o(\mu_1,\mu_2,t)\text{cos}\theta\text{sin}\frac{\beta}{2}
\end{split}
\label{rhom}
\end{equation}
where two imbalance functions ${\cal F}_e$ and ${\cal F}_o$ are introduced to quantify the nonequilibrium contribution
\begin{equation}
\begin{split}
& K_v = 1-2\left[f_v^{\mu_1}\left(1-f_v^{\mu_2}\right)+f_v^{\mu_2}\left(1-f_v^{\mu_1}\right)\right]\text{sin}^2\left(\frac{f}{J}\text{sin}\frac{\beta}{2}\right)\\[0.2cm]
& \mathcal{F}_e\left(\mu_1,\mu_2,t\right) = \sum_{\{m_e\}}(-4)^{\frac{\text{c}(m_e)}{2}}\text{sin}^{\text{c}(m_e)}\left(\frac{2f}{J}\text{sin}\frac{\beta}{2}\right)\prod_{v\in\bar{m}_e}K_v\prod_{q\in m_e}\left(f_q^{\mu_1}-f_q^{\mu_2}\right)\\[0.2cm]
& \mathcal{F}_o\left(\mu_1,\mu_2,t\right) = \sum_{\{m_o\}}(-4)^{\frac{\text{c}(m_o)-1}{2}}\text{sin}^{\text{c}(m_o)}\left(\frac{2f}{J}\text{sin}\frac{\beta}{2}\right)\prod_{v\in\bar{m}_o}K_v\prod_{q\in m_o}\left(f_q^{\mu_1}-f_q^{\mu_2}\right)
\end{split}
\label{amb}
\end{equation}
\end{widetext}
and $m=(m_1,m_2,\cdots,m_s),\ s=1,2,\cdots,N$ is a non-empty subset of the set $r=(1,2,\cdots,N)$; $\text{c}(m)$ is the number of elements of subset $m$. $m_e,\ m_o$ correspond to the $m$'s with even and odd numbers of elements, respectively. $\bar{m}=r-m$.
$f_n^{\mu_i}=\left[\text{exp}\left(\frac{\hbar\nu_n-\mu_i}{k_B T}\right)+1\right]^{-1},\ \hbar\nu_n = \varepsilon + 4t\text{cos}\left(2n\pi/N\right)$. So far, we have obtained the {\it exact} dynamics of spin qubits beyond the Markovian and weak qubit-bath coupling approximations. In the forthcoming discussion, we will focus on the nonequilibrium effects reflected by the imbalance functions ${\cal F}_e$ and ${\cal F}_o$ which are governed by the chemical potential difference of spin baths.

\section{Recurrence from relaxation of spin baths}
Based on our exact dynamics obtained above, we are able to explicitly explore the non-Markovian process beyond Markovian approximation without memory. 
This recently attracted much attention in the study of dissipative quantum dynamics because the experimental measurements revealed the non-Markovian noise from the nonexponential decay of echo signal \cite{Dial13}.

\begin{figure*}[t]
 \centering
 $\begin{array}{c}
   \includegraphics[scale=0.61]{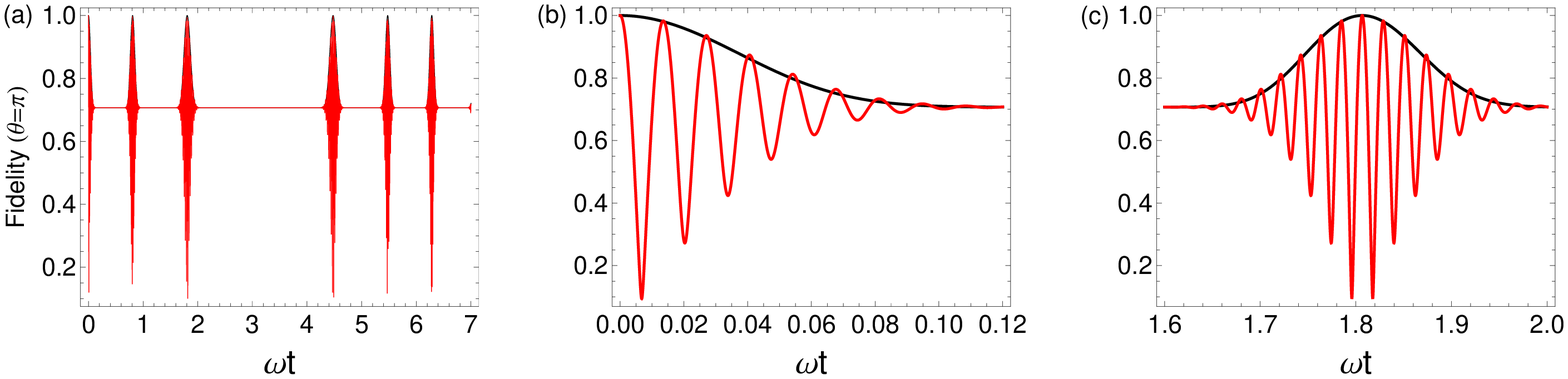}\\
   \includegraphics[scale=0.6]{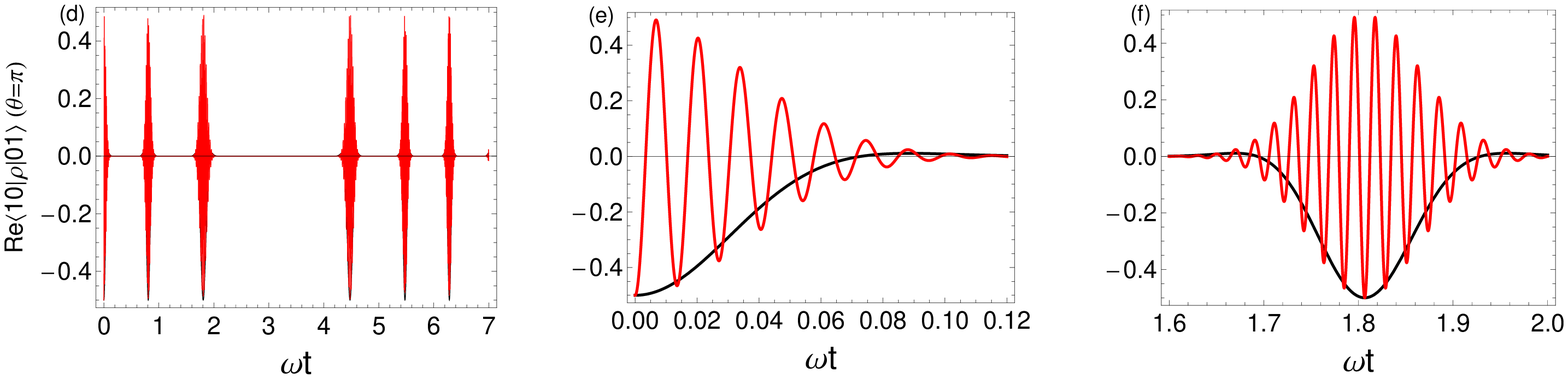}\\
   \includegraphics[scale=0.61]{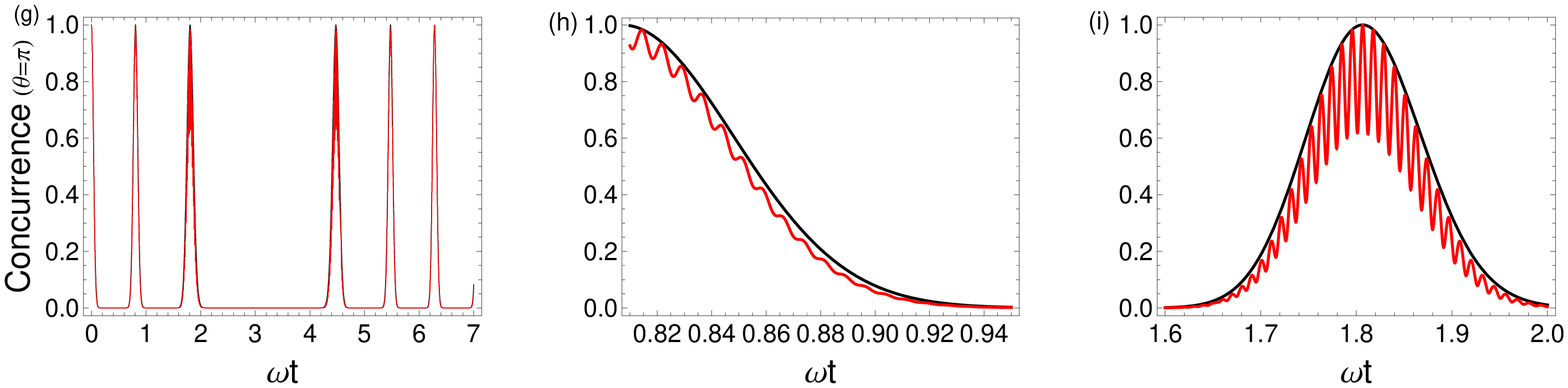}
  \end{array}$
\caption{(Color online) Dynamics of (top) fidelity of quantum state $|\Psi_{\pi}(0)\rangle=\frac{1}{\sqrt{2}}(|1\rangle\otimes |0\rangle-|0\rangle\otimes |1\rangle)$, (middle) the real part of coherence $\text{Re}\langle 1,0|\rho_s|0,1\rangle$ and (bottom) concurrence of the spin qubits; $\omega\equiv 4J/\hbar$. The red and black lines are for $\mu_1=1.7,\ \mu_2=0.5$ (far-from-equilibrium) and $\mu_1\simeq\mu_2=0.5$ (equilibrium), respectively. Other parameters are $\varepsilon=1$, $t=0.2$, $k_B T=0.1$, $N=100$ and $f=8J$.}
\label{f2}
\end{figure*}

To measure the preservation of the quantum state, one can use the {\it fidelity} of the state defined as $F(t)=\sqrt{\langle\Psi(0)|\rho_s(t)|\Psi(0)\rangle}$ if the system is initially in pure ensemble. Equivalently, the fidelity can be written as
\begin{equation}
\begin{split}
F(t) = \sqrt{\frac{1}{2}+\text{cos}\theta\text{Re}\langle 1,0|\rho_s|0,1\rangle+\text{sin}\theta\text{Im}\langle 1,0|\rho_s|0,1\rangle}
\end{split}
\label{fidelity}
\end{equation}
which shows a strong correlation to the quantum coherence. According to recent experiments \cite{Yacoby11}, the spin qubits are initially prepared at a singlet state $|\Psi_{\pi}(0)\rangle=\frac{1}{\sqrt{2}}(|1\rangle\otimes|0\rangle-|0\rangle\otimes|1\rangle),\ (\theta=\pi)$ in the regime of strong qubit-nuclear interactions, whose dynamical behaviors are illustrated in Fig.\ref{f2}. As is shown, the preservation of quantum state and revival of coherence are perfectly elucidated. Here we only show the behavior of the real part of coherence because it contributes to the preservation of state, based on Eq.(\ref{fidelity}). Physically the recovery of quantum coherence can be attributed to the non-Markovian effect \cite{Wang1993_CPL,Wang1995_PRL}, since the timescale of correlations in the environment is comparable to that of the system and the phase correlation of the system has a high chance to reconstruct. This implies the nonlocal correlation of the system in time domain and consequently the information of initial state is memorized which results in the recurrence of coherence.
The coherence always shows a monotonic decay until reaching the stationary value. Furthermore it is worth noticing that the system shows a perfect preservation of quantum state with fidelity of 100\%. This feature can be alternatively understood from the coherence dynamics $\text{Re}\langle 1,0|\rho_s|0,1\rangle$, shown in Fig.\ref{f2}(d), in which the magnitude of coherence periodically recovers to its initially value. The type of state protection we discuss here is completely different from what has been achieved by dynamical decoupling and other control methods \cite{Uhrig07,Sarma07,Sarma12,Sarma14,Sarma15}. It is instead an intrinsic entanglement induced by slow relaxation of the baths leading to the non-Markovian process. This means no need to artificially control over the spin-spin interactions to combat decoherence.

Moreover, no global decay occurs for the local peaks of either coherence or fidelity with initially being engineered at the Bell state, as compared to that in the case $|\Psi_{\frac{\pi}{2}}(0)\rangle=\frac{1}{\sqrt{2}}(|1\rangle\otimes|0\rangle-i|0\rangle\otimes|1\rangle),\ (\theta=\frac{\pi}{2})$ as shown in Supplementary Materials (SM) where such global decay does exist. This is because of the conservation of the total angular momentum in the case $\theta=\pi$: $[L_z,H]=0$ ($|\Psi_{\pi}(0)\rangle$ is the eigenstate of total angular momentum $L_z=\sigma_1^z+\sigma_2^z$). 

To show the generality of the non-Markovian effect, we also perform the dynamics of fidelity and coherence as the spin qubits relax from another state $|\Psi_{\frac{\pi}{2}}(0)\rangle=\frac{1}{\sqrt{2}}(|1\rangle\otimes|0\rangle-i|0\rangle\otimes|1\rangle),\ (\theta=\frac{\pi}{2})$, whose dynamics is illustrated in SM. As is shown, the memory effect arising from the non-Markovian process always shows up, irrespect to which initial state is prepared. The generality of such non-Markovian effect is further demonstrated by letting the qubits to relax from other states, e.g., $|\Psi_{\frac{\pi}{4}}(0)\rangle=\frac{1}{\sqrt{2}}(|1\rangle\otimes|0\rangle+e^{-i\pi/4}|0\rangle\otimes|1\rangle)$ as also shown in SM.

\begin{figure*}
 \centering
 $\begin{array}{cc}
   \includegraphics[scale=0.49]{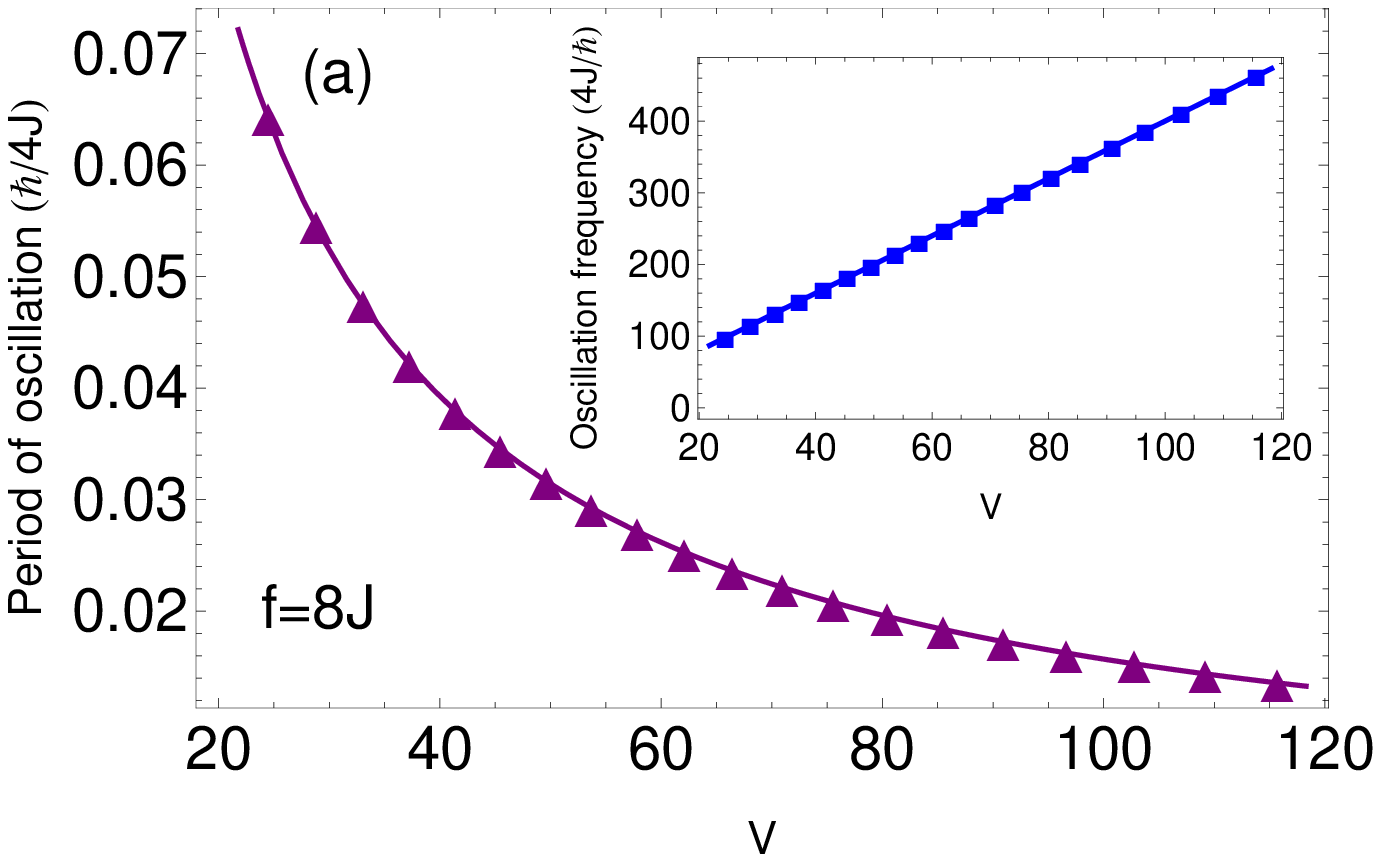}
  &\includegraphics[scale=0.48]{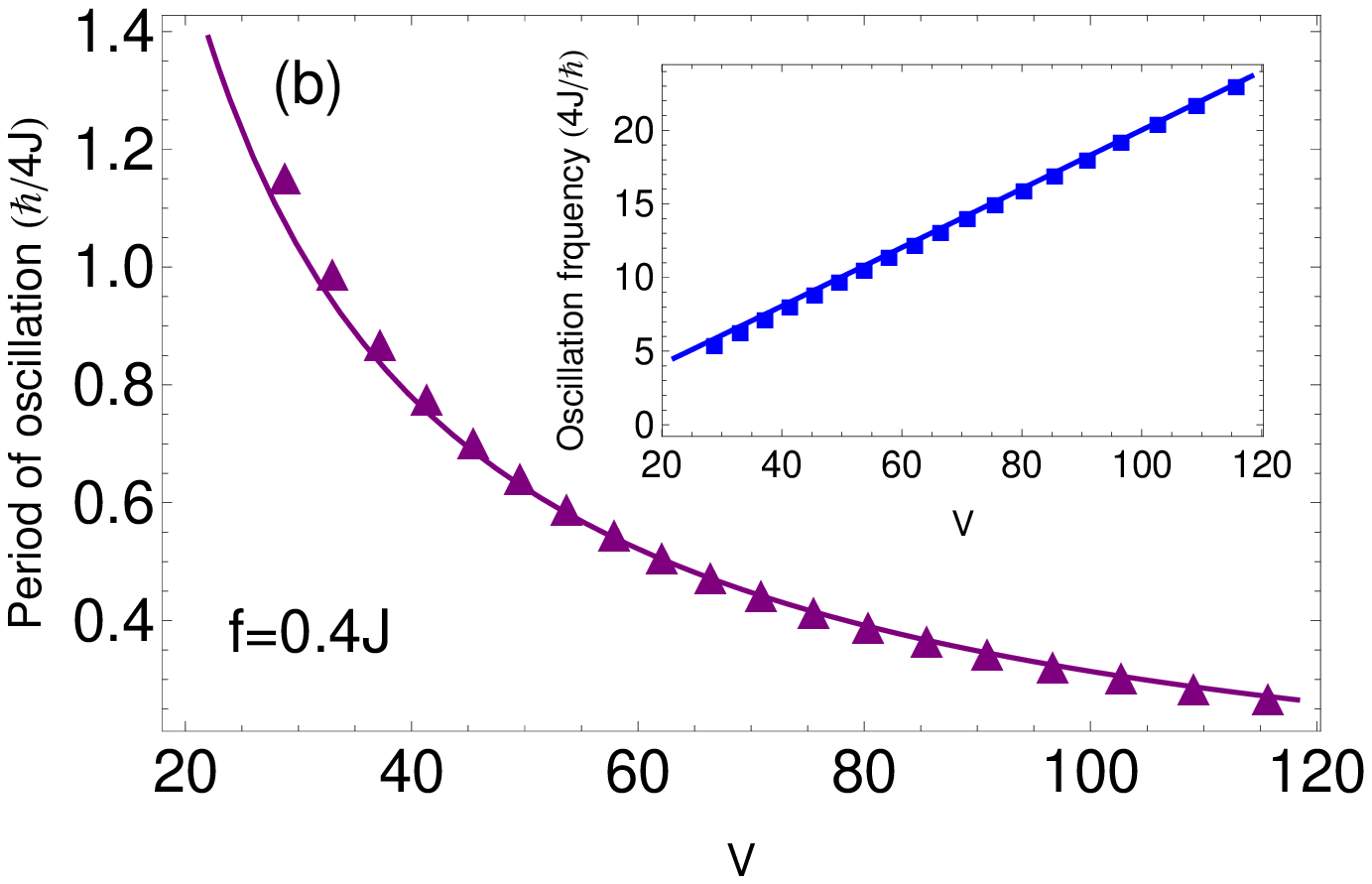}\\
   \includegraphics[scale=0.49]{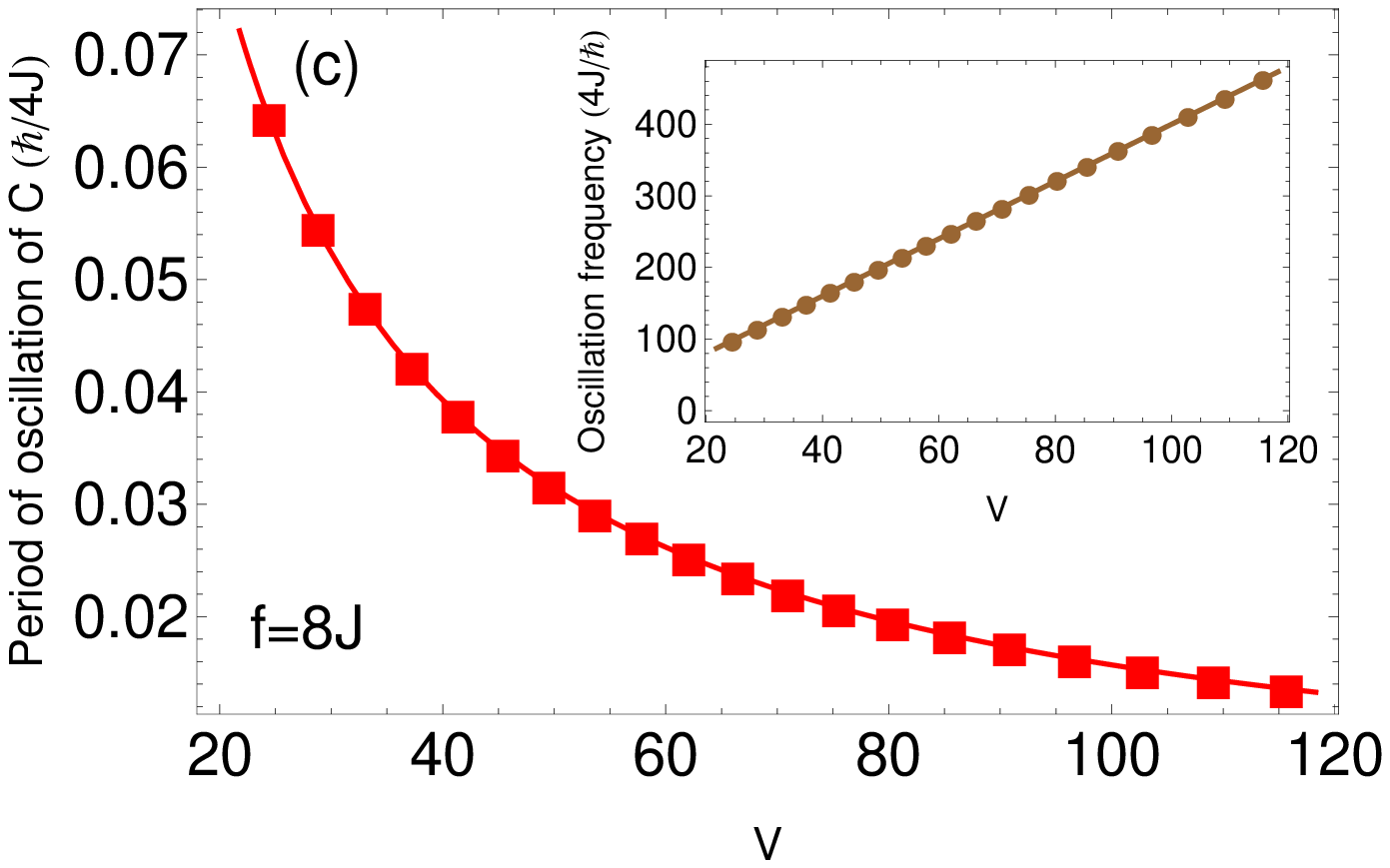}
  &\includegraphics[scale=0.48]{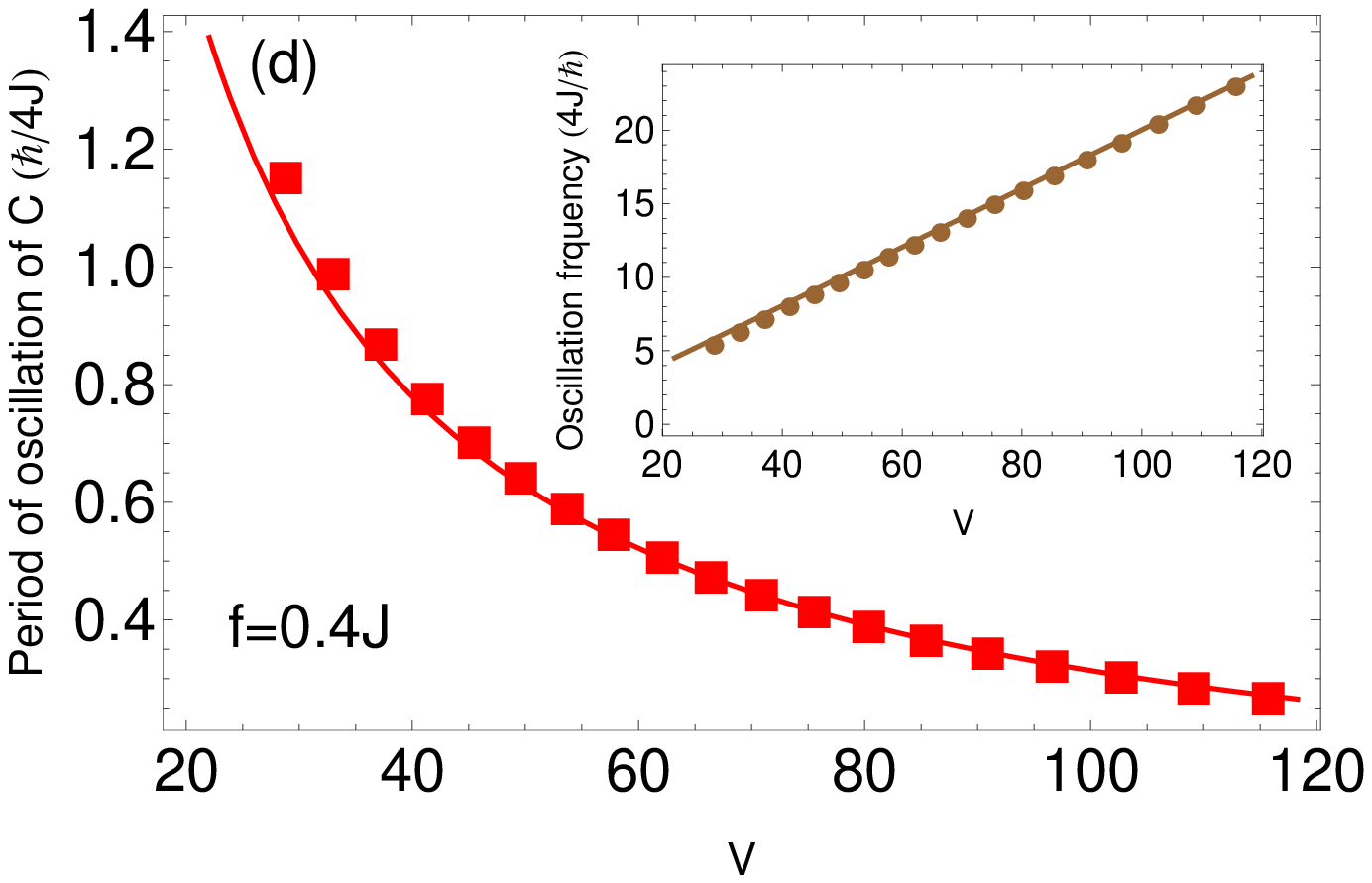}\\
  \end{array}$
\caption{(Color online) (Large) Oscillation period and (Small) oscillation frequency for (Top) coherence and (Bottom) concurrence vary as a function of effective voltage $V$, for (a,c) strong ($f=8J$) and (b,d) weak ($f=0.4J$) qubit-nuclear interactions; The triangle and circle markers are for the numerical calculations of oscillation period and frequency, respectively; For oscillation period, the purple triangle and red square markers correspond to the coherence and concurrence, respectively. All the smooth curves are obtained from the analytical result in Eq.(\ref{Omega}). Other parameters are $\varepsilon=1$, $t=0.2$, $k_B T=0.1$ and $N=100$.}
\label{f3}
\end{figure*}

\section{Enhancement of dynamical coherence from detailed-balance-breaking}

Now let us turn to the nonequilibrium effect with the detailed-balance-breaking induced by the gradient of chemical potential from nuclear spins. From our analytical solution of the density matrix we know that the nonequilibrium contribution can be quantified by the factor $\prod(f_q^{\mu_1}-f_q^{\mu_2})$ in the imbalance functions ${\cal F}_e$ and ${\cal F}_o$ which vanish under the time-reversal-protection. The time evolution shown in Fig.\ref{f2} and the figure in SM illustrate that the nonequilibriumness produces the rapid oscillations, through the comparison between the red and black lines for far-from-equilibrium and equilibrium regimes, respectively. This reveals that the far-from-equilibrium is essential for observing the coherent oscillation of spin qubits in the experiments. Later we will discuss the experimental implementation of such effect based on quantum simulation.

To understand and explain the oscillation feature produced by the detailed-balance-breaking, we need to introduce the spin current between the spin qubits which provides a measure of the degree of deviation from equilibrium \cite{Zhang14,Qian2006_JPCB,Qian1979_book}. The current conservation gives $\frac{d}{dt}\sigma_2^{\dagger}\sigma_2=\hat{I}_{1\rightarrow 2}-\hat{I}_{2\rightarrow bath}$ where $\hat{I}_{1\rightarrow 2}$ denotes the spin current operator whose expectation $I_{1\rightarrow 2}=\langle \hat{I}_{1\rightarrow 2}\rangle$ gives the spin current. In our model the current from the system to spin environment vanishes because of $[\sigma_2^{\dagger}\sigma_2,\sigma_i^z S_{n,i}^z]=0;\ i=1,2$. 
Thus the spin current reads
\begin{equation}
\begin{split}
I_{1\rightarrow 2} = \frac{d}{dt}\langle \sigma_2^{\dagger}\sigma_2\rangle = \frac{4J}{\hbar}\text{Im}\langle 1,0|\rho_s|0,1\rangle
\end{split}
\label{current}
\end{equation}
based on the Heisenberg's equation. This coincides with the form for curl quantum flux in our former work \cite{Zhang14}, in that microscopically the current strongly correlates to the curl quantum flux, vanishing under detailed-balance at steady state. However this current does not necessarily vanish during the non-equilibrium relaxation to the steady state. Even at the steady state, the current is not necessarily zero due to the energy or information input from or output to the environments. As is shown by Eq.(\ref{current}), the spin current governed by the nonequilibriumness generates a fast oscillation of coherence, furthermore the fidelity of quantum state. This can be understood as follow: {\it the rapid oscillation of spin current induced by chemical potential imbalance (nonequilibriumness) leads to the back and forth motion of spin waves between qubtis, which results in the fast oscillations of coherence}. For an analogy, this is in the similar spirit of limit cycle behavior in classical stochastic processes \cite{Gang93,Rappel94}, driven by the curl flux breaking the detailed-balance at steady state, where a robust oscillation network can be observed \cite{Wolynes07,Wang08,Li12}.

To further understand the nonequilibrium effect, we consider a certain limit where the nuclear spin environments evolve sufficiently slowly so that they can be well approximated by quasistatic ensembles \cite{Wang1993_CPL,Wang1995_PRL}, namely $\frac{d S_{v,i}^z}{dt}\simeq 0$ on the typical timescales of electron spin dynamics, around a microsecond or less \cite{Barnes12,Cywinski09prl,Cywinski09prb}. Then the entire system can be approximately described by product state $\rho(t)\simeq\rho_s(t)\otimes\rho_{nc}^{\mu_1}\otimes\rho_{nc}^{\mu_2}$. By adopting the Heisenberg-Langevin theory \cite{Scully1997_book} one can obtain the coherent oscillation: $\text{Im}\langle\sigma_1^+\sigma_2^-\rangle\propto\text{cos}(\Omega_{\text{coh}}t+\phi)$ and $\text{Re}\langle\sigma_1^+\sigma_2^-\rangle\propto\frac{1}{\Omega_{\text{coh}}}\text{sin}(\Omega_{\text{coh}}t+\phi)$ where the oscillation frequency reads
\begin{equation}
\begin{split}
\Omega_{\text{coh}} = \frac{4J}{\hbar}\sqrt{1+\frac{f^2V^2}{4J^2}},\ \ V = 2\sum_{q=1}^N\left(f_q^{\mu_1}-f_q^{\mu_2}\right)
\end{split}
\label{Omega}
\end{equation}
$V$ serves as an {\it effective voltage} vanishing under detailed-balance and it provides one type of quantification for the chemical potential imbalance. Eq.(\ref{Omega}) uncovers the analytical relation between the coherent oscillation and the nonequilibriumness. It explicitly demonstrates that the detailed-balance-breaking is intrinsically responsible for the rapid oscillation of coherence and subsequently the fidelity of quantum state, which confirms our argument above. To verify the validity of our formula for $\Omega_{\text{coh}}$, we perform a numerical calculation of the oscillation period and frequency with respect to voltage $V$, illustrated in Fig.\ref{f3}, which shows a perfect agreement with our analytical formula Eq.(\ref{Omega}).

It is worth pointing out that the case discussed above with static nuclear spin environment can be alternatively described by a back-of-the-envelope model with the Hamiltonian $H_1=-J(\sigma_1^x\sigma_2^x+\sigma_1^y\sigma_2^y)+\sigma_1^z(B+\Delta B)+\sigma_2^z B$. This is due to the static magnetic fields $B+\Delta B$ and $B$ produced by nuclear spins. The static magnetic field does not break the time reversal whereas our model studied before does, as indicated by the damping oscillation. Some algebra gives rise to the Rabi frequency $\Omega_1=\frac{4J}{\hbar}\sqrt{1+\frac{\Delta B^2}{4J^2}}$ which shows the increase of oscillation frequency by the magnetic field gradient $\Delta B$. As $\Delta B\propto \sum(\langle S_{n,1}^z\rangle-\langle S_{n,2}^z\rangle)$, the results in Eq.(\ref{Omega}) can then be recovered. This coincidence further demonstrates the fact that the rapid oscillation of quantum coherence is attributed to the nonequilibriumness produced by the inhomogenous charge density of the nuclei environment.


\section{Quantum entanglement enhanced by detailed-balance-breaking}
The quantum nature usually is not only reflected by coherence, but is also revealed by quantum entanglement. The latter one takes the advantage over the former one because of its basis-independence. In other words, the measure of quantum coherence depends on the choice of basis, so that it can be observed under some specific basis while it may vanish by switching to other basis. Although  the entanglement can be described by entanglement entropy in an elegant way for closed quantum systems, how to quantitatively measure the entanglement for the open quantum systems is still a challenging issue. Despite of this, several quantifications have been proposed,  i.e., negativity \cite{Werner2002_PRA} and concurrence \cite{Wootters1998_PRL}, each of which however, has its own limitation. Here we will quantify the entanglement by exploring the concurrence of the spin qubits, taking the advantage of the spin-$\frac{1}{2}$ feature.

To obtain the concurrence which quantifies the quantum entanglement of the spin-qubit system, the spin-flipped operation $\tilde{\rho}=(\sigma^y\otimes\sigma^y)\rho^*(\sigma^y\otimes\sigma^y)$ to the system must be carried out at first and
\begin{equation}
\begin{split}
\tilde{\rho}(t) = \begin{pmatrix}
                   \rho_{01,01}(t) & \rho_{10,01}(t)\\[0.15cm]
                   \rho_{10,01}^*(t) & \rho_{10,10}(t)\\
                  \end{pmatrix}
\end{split}
\label{rhotilde}
\end{equation}
where $\rho^*$ is the complex conjugate of $\rho$. Thus the non-Hermitian matrix $R=\rho\tilde{\rho}$ can be obtained
\begin{equation}
\begin{split}
R = \begin{pmatrix}
     \left(\rho_{10,10}\rho_{01,01}+|\rho_{10,01}|^2\right) & 2\rho_{10,10}\rho_{10,01}\\[0.15cm]
     2\rho_{01,01}\rho_{10,01}^* & \left(\rho_{10,10}\rho_{01,01}+|\rho_{10,01}|^2\right)\\
    \end{pmatrix}
\end{split}
\label{R}
\end{equation}
whose eigenvalues are denoted as $\lambda_i,\ i=1,2,3,4$ in descending order. Obviously in our setup $\lambda_3=\lambda_4=0$ and subsequently the concurrence is ${\cal C}(\rho) = \text{max}\left(0,\sqrt{\lambda_1}-\sqrt{\lambda_2}-\sqrt{\lambda_3}-\sqrt{\lambda_4}\right)$ which gives rise to
\begin{equation}
\begin{split}
{\cal C}(\rho) & = \sqrt{\rho_{10,10}\rho_{01,01}}+|\rho_{10,01}|-\big|\sqrt{\rho_{10,10}\rho_{01,01}}-|\rho_{10,01}|\big|\\[0.15cm]
               & = 2|\rho_{10,01}| 
\end{split}
\label{conc}
\end{equation}
The positivity of the density matrix requires that the eigenvalues $p_1,\ p_2$ must be non-negative. Thus $p_1p_2=\rho_{10,10}\rho_{01,01}-|\rho_{10,01}|^2\ge 0$, which gives the result in the 2nd line in Eq.(\ref{conc}).

Eq.(\ref{conc}) clearly shows that the entanglement will die when the coherence is destroyed. This on the other hand, reveals the intrinsic correlation between the quantum coherence and entanglement. The concurrence can be properly employed to quantify the entanglement ${\cal E}$ due to the fact that it is monotonically increasing with respect to ${\cal E}$ for $0\le {\cal C}\le 1$. Fig.\ref{f2}(g,h,i) illustrates the dynamics of concurrence based on Eq.(\ref{conc}), which shows the similar rapid oscillation generated by nonequilibriumness as that occurs in coherence dynamics. This demonstrates that the entanglement between qubits can be enhanced by detailed-balance-breaking. This further manifests the fact in a solid manner that the far-from-equilibrium regime can not only promote the steady-state coherence \cite{Zhang14,Zhang15pccp}, but can also improve the quantum nature in terms of entanglement in the dynamical processes. On the other hand, the non-Markovian effect is also displayed clearly by the recovery of entanglement shown in Fig.\ref{f2}(g,h,i). To further explore the nonequilibrium contribution to the oscillation period which can be measured through the  experiments, we find that from Eq.(\ref{conc}) and the procedures for obtaining the oscillation frequency of coherence, the quantitative relation between voltage and the oscillation period of entanglement is of the form
\begin{equation}
\begin{split}
T_{\text{concur}} = \frac{\pi\hbar}{2J}\left(1+\frac{f^2V^2}{4J^2}\right)^{-\frac{1}{2}}
\end{split}
\label{Tc}
\end{equation}
which shares the same form as Eq.(\ref{Omega}), owing to Eq.(\ref{conc}). Eq.(\ref{Tc}) predicts the rapid coherent oscillation of entanglement in far-from-equilibrium regime. The numerical calculations confirm our analytical formula for oscillation period, as illustrated by the red markers and lines in Fig.\ref{f3}(c) and (d).

\section{Experimental implementation for simulating spin-qubit dynamics}

\begin{figure*}
 $\begin{array}{cc}
   \includegraphics[scale=0.45]{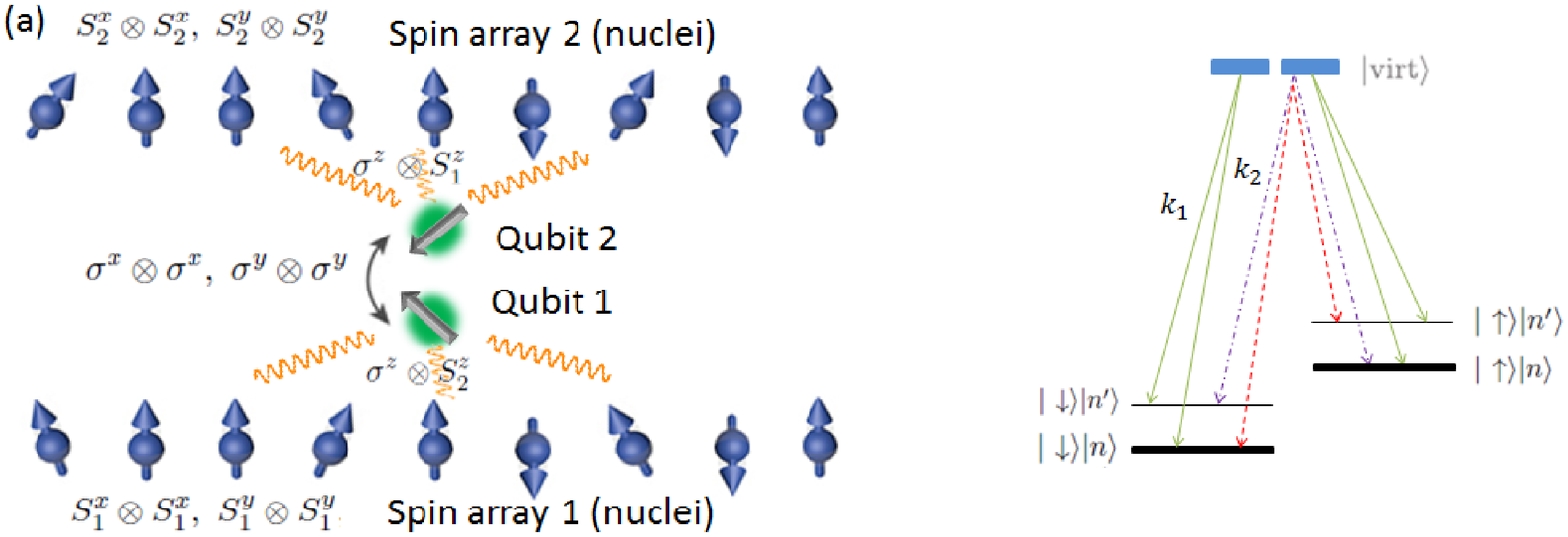}\\[0.4cm]
   \includegraphics[scale=0.6]{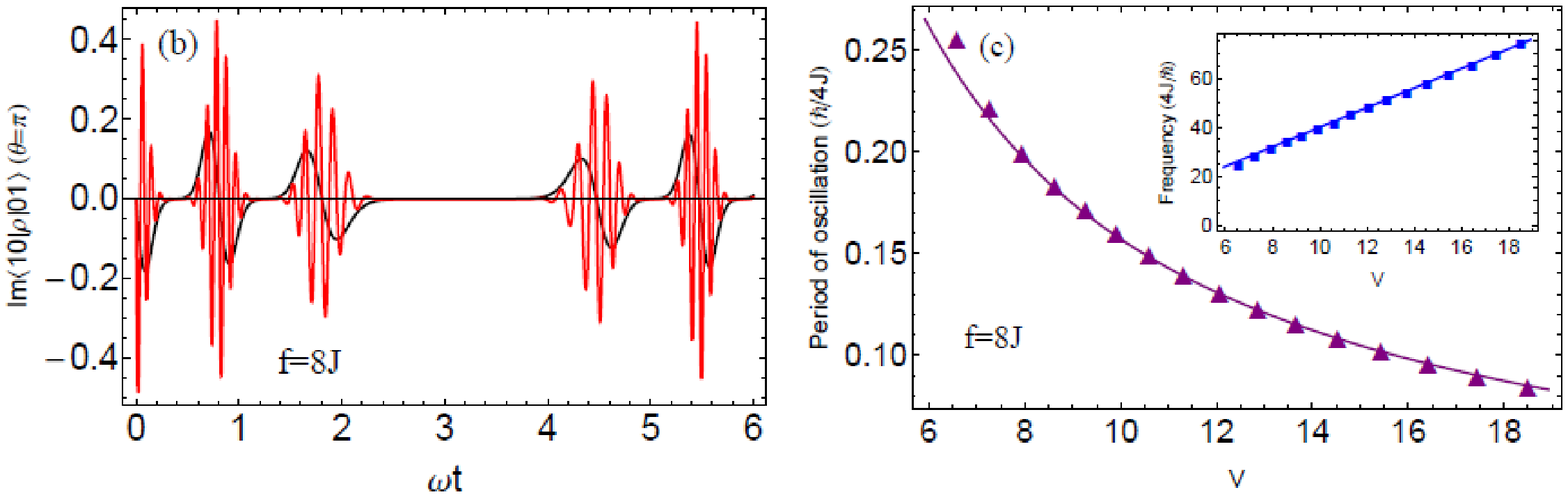}
  \end{array}$
\caption{(Color online) (a) Schematics of (Left) our setup and (Right) the stimulated two-photon Raman transition for realizing (Red and blue) $\sigma^x\otimes\sigma^x+\sigma^y\otimes\sigma^y$ and (Orange) $\sigma^z\otimes\sigma^z$ interactions; (b) Dynamics of coherence $\text{Im}\langle 1,0|\rho_s|0,1\rangle$ with the initial condition $|\Psi(0)\rangle=\frac{1}{\sqrt{2}}(|1\rangle\otimes |0\rangle-|0\rangle\otimes |1\rangle)$ and (c) the period (small for frequency) of coherent oscillation as a function of voltage $V$.  The red and black lines in (a) are for $\mu_1=2.22\text{GHz},\ \mu_2=0.65\text{GHz}$ (far-from-equilibrium) and $\mu_1\simeq\mu_2=0.65\text{GHz}$ (equilibrium), respectively; The triangle and square markers are for the numerical calculations of oscillation period and frequency, respectively. The smooth curves are obtained from Eq.(\ref{Omega}). Other parameters are $\varepsilon\simeq 1.31\text{GHz},\ t\simeq 0.26\text{GHz},\ T\simeq 1\text{mK},\ J\simeq 0.22\text{GHz},\ f\simeq 1.76\text{GHz}$ and $N=16$.}
\label{ions}
\end{figure*}

In order to observe the non-Markovian and nonequilibrium effects investigated in this work, ultracold trapped ions seems to be a good candidate to engineer our setup, since they are recently used for the spin chain simulations \cite{Wunderlich09,Islam11,Duan11,Roos12}. 

The ions are usually confined by a linear radio-frequency (RF) trap with three-dimensional electrodes, which subsequently produces the internal levels of ions. The spin-$\frac{1}{2}$ structure is realized by choosing two nearly degenerate sublevels of the ground state, splitted by Zeeman field. On the other hand, the ions reside on individual lattice sites due to the strong Coulomb repulsion, which still leads to the common motional modes of ions described by phonons. Such phonons allow long-range interactions to be mediated between spins associated with the ions. To engineer the long-range spin-spin interactions, two detuning laser beams with different frequencies $\omega_1,\ \omega_2$ are required to perform stimulated two-photon Raman transition between the sublevels, through a third level with higher energy, as shown in Fig.\ref{ions}(right). Here the two particular processes are crucial: (a) the transition $|\uparrow\rangle|n'\rangle\Leftrightarrow|\downarrow\rangle|n\rangle\ (n'\neq n)$ with two laser beams detuned by the frequency difference between these two states; (b) the transitions $|\uparrow\rangle|n'\rangle\Leftrightarrow|\uparrow\rangle|n\rangle\ (n'\neq n)$ and $|\downarrow\rangle|n'\rangle\Leftrightarrow|\downarrow\rangle|n\rangle\ (n'\neq n)$ with the two beams detuned by approximately the
frequency of a motional mode. Here $n',n$ denote the energy levels of ions. It can be directly shown that (a) generates the transversed long-range interactions between spins, $\sigma^x\otimes\sigma^x$ and $\sigma^y\otimes\sigma^y$ depending on the polarization of electric field, while (b) generates the Ising type of interaction, $\sigma^z\otimes\sigma^z$ \cite{Cirac04,Schaetz12}.

The array of nuclear spins can be implemented by an array of $N$ trapped ions, showing a dipolar decay of spin-spin interaction $J_{ij}\sim \frac{1}{|i-j|^3}$. 
In order to realize the XY couplings, the spin array can be manipulated to interact with detuning laser beams in terms of process (a) above. We now prepare other two ions trapped by linear RF trap, with the same splitting between the sublevels of ground state as that in arrays of nuclear spins. The same technique can be employed as before to produce the transversed spin-spin interaction between these two ions. To engineer the interaction between spin arrays and the two ions, the process (b) can be used to generate the individual coupling of ion 1(2) to each ion in the spin array 1(2), by choosing the $\hat{z}$-polarization of the electric field in the laser beams, illustrated by Fig.\ref{ions}(left). This indicates that $2N$ pairs of detuning beams with $\hat{z}$-polarization are demanded in total, to simulate the qubit-nuclear interactions.

We use 16 trapped ions to simulate each nuclear spin environment, by taking into account the conditions in recent experiments \cite{Islam11,Duan11,Roos12}. The two arrays of ions can be prepared initially with different Fermi energies and the two ions (qubits) are engineered at singlet state $|\Psi(0)\rangle=\frac{1}{\sqrt{2}}\left(|1\rangle\otimes |0\rangle-|0\rangle\otimes |1\rangle\right)$. By choosing the parameters $\varepsilon\simeq 1.31\text{GHz},\ t\simeq 0.26\text{GHz},\ T\simeq 1\text{mK},\ J\simeq 0.22\text{GHz},\ f\simeq 1.76\text{GHz},\ \mu_2\simeq 0.65\text{GHz}$ and $N=16$ based on the experimental feasibility \cite{Schaetz12,Roos12}, Fig.\ref{ions} shows the dynamics of coherence and also the behavior of frequency of coherent oscillation with respect to effective voltage as introduced before. The non-Markovian effect is clearly shown by the revival of the coherence in Fig.\ref{ions}(b). Moreover, it also shows the nonequilibrium effect (detailed-balance-breaking) which generates the fast coherent oscillation, through the comparison between the red and black curves. The dependence of oscillation period (frequency) is further illustrated in Fig.\ref{ions}(c), which is measurable in the proposed experiment.

\section{Discussion and conclusion}
In summary, we exactly and analytically solved the dynamics of spin qubits surrounded by the charge noise produced by nuclear spins. We found that the detailed-balance-breaking leads to a new phase of rapid coherent oscillation in spin dynamics, which was lacking in the conventional ensemble of nuclear spin bath preserving the detailed-balance, i.e., Overhauser noise. The analytical relationship between oscillation frequency of coherence \& entanglement and effective voltage was further obtained. It suggests a quantitative measure of the nonequilibrium effect, which would be accessible in experiments. On the other hand, our results have the advantage over previous studies in purely numerical manner for describing the non-Markovian process. Thus we can predict the recovery of coherence and the subsequent quantum-state preservation in a general scenario. These novel effects we predict, especially the oscillations of coherence and entanglement arising from detailed-balance-breaking, can be observed in ultracold trapped ions we proposed in details in the spirit of quantum simulation in the laboratory.

\section{Acknowledgements}
We acknowledge the support from the grant NSF-PHY-76066.





\end{document}